\documentclass{elsart4}
\usepackage{graphicx}
\usepackage{amsmath}
\usepackage{amssymb}

\begin{document}

\begin{frontmatter}

\title{
Zero-energy edge states and 
chiral symmetry breaking 
at edges of graphite sheets}

\author[address1]{Shinsei~Ryu\thanksref{thank1}},
and
\author[address1]{Yasuhiro~Hatsugai} 

\address[address1]{Department of Applied Physics, University of Tokyo, 7-3-1 Hongo Bunkyo-ku, 
Tokyo 113-8656, Japan}

\thanks[thank1]{
Corresponding author. 
E-mail: ryuu@pothos.t.u-tokyo.ac.jp}

\begin{abstract}
Two-dimensional graphite sheets with a certain type of edges
are known to support boundary states localized near the edges.
Forming a flat band with
a sharp peak in the density of states at the Fermi energy,
they can trigger a magnetic instability
or a distortion of the lattice
in the presence of electron-electron or electron-phonon interactions.
We shall discuss
a relationship between chiral symmetry,
which is the origin of the zero-energy edge states,
and several types of induced orders such as 
spin density waves or lattice distortions.
We also 
investigate electron correlation effects 
on the edge states 
for a wrapped quasi one-dimensional geometry,
i.e., carbon nanotube, 
by means of the renormalization group and
the open boundary bosonization.
\end{abstract}

\end{frontmatter}


\section{Introduction}

A fundamental aspect of two-dimensional (2D) graphite sheet 
is its intriguing band structure
caused by the honeycomb network formed by $\sigma$-bonds ($sp^{2}$)
on which $\pi$ electrons are 
performing quantum mechanical hoppings.
The nontriviality of the band can be
quantitatively (and mathematically) characterized 
in terms of the Berry (Zak) phase,
which is a phase degree of freedom in the quantum mechanical systems.
The importance of the underlying gauge structure (the Berry phase)
has also been recognized in the context 
of the integer quantum Hall effect
or a theory of macroscopic polarizations\cite{kingsmith93}.

A direct consequence of the nontriviality of the band
is the appearance of edge states 
which are localized near the edges
when the system is truncated with a certain type of edges.
Indeed,  Fujita et al. discovered 
peculiar edge states forming a flat band at zero-energy (the Fermi energy)
in graphite sheets with zigzag or bearded (Klein) edges.
\cite{fujita96}

In the present paper,
we consider edge states physics
caused by electron correlation effects or 
coupling with lattice distortions on the honeycomb lattice 
with several types of edges.
One of our main messages is that 
patters of order parameters 
[spin density waves (SDW), lattice distortions, etc.]
which can be induced in the presence of interactions
are dictated by a symmetry called chiral symmetry
that is responsible for the existence of the zero-energy edge states (ZES's).
Our argument is generic
and can be applicable to other situations
such as coexistence of time-reversal symmetry breaking
order parameters at (110) surfaces of $d_{x^{2}-y^{2}}$-wave superconductors.
\cite{ryu02,ryu02-02}

As an example,
we shall discuss 
coupling between $\pi$ electrons and lattice distortions
in terms of the Su-Schrieffer-Heeger (SSH) model.
A possibility of a SDW order within the mean field theory (MFT) 
for the Hubbard model defined on the honeycomb lattice with edges
is also considered.
We also give detailed discussions
on electron correlation effects when a graphite sheet 
are wrapped into quasi-1D geometry,
i.e., carbon nanotube (CNT),
where correlation effects are much more pronounced
due to its low-dimensionality.
The emergence of magnetic moments at edges 
is attractive since 
it opens up a possibility
to have a new magnetic material made
of exclusively light elements,
without $d$- or $f$- electrons.
\cite{kusakabe02}

\section{Origin of zero-energy edge states and a Jahn-Teller-like argument}

\begin{figure}[h]
\begin{center}
\includegraphics[width=7cm,clip]{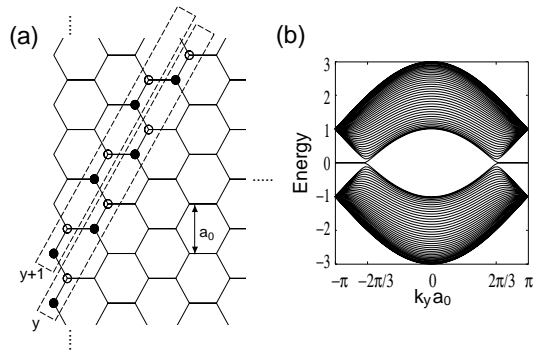}
\caption{
\label{lattice}
(a) A $(N,-N)$ CNT with a zigzag edge.
Unit cells are shown by broken lines.
(b)
The corresponding energy spectrum.
}
\end{center}
\end{figure}

Typical examples of a truncated graphite sheet
that supports edge states are 
those terminated with zigzag edges
in which 
edge states appear for
$+2\pi /3 < |k_{y}a_{0}| \le +\pi$,
where $k_{y}$ is the wave number along the edges
and $a_{0}$ the lattice constant (Fig. \ref{lattice}).
On the other hand,
a graphite sheet with armchair edges 
does not support edge states.
The band structure for the zigzag case is nontrivial
as its Zak phase is equal to $\pi$ for $k_{y}$ specified above.
In the present case,
nontriviality is also conveniently 
characterized in terms of a topological object,
a set of loops in a parameter space,
from which we can infer the existence of edges states.
\cite{ryu02}

Although nontriviality of the band implies the existence of 
edge modes, 
it does not assure that edges states are located at zero-energy.
Chiral symmetry,
which means 
there is no matrix element
connecting two sublattices in the present case,
is another ingredient that dictates 
edge modes to be at zero-energy.
Similar ZES's
are also known for other systems 
such as surface states of $d_{x^{2}-y^{2}}$-wave superconductors,
and a states localized at a soliton in polyacetylene.
All such examples can be understood
in terms of the topology of bands and chiral symmetry
in a unified fashion.
\cite{ryu02}

As the edge states form a flat band
(Fig. \ref{lattice}), 
giving rise to a sharp peak in the density of states (DOS) at the Fermi energy,
a natural question is then what can happen 
in the presence of electron-electron or electron-phonon interactions.
Once we understand the origin of ZES's,
it is possible to have some insights as to what kind of orders can occur
as follows.
Within the mean field picture,
the band structure near the edges
can be effectively modified when there are interactions
to lift the degeneracy and hence lower the ground state energy.
However, as ZES's are ``protected'' by chiral symmetry,
such modifications should be accompanied 
with the breaking of chiral symmetry near the boundaries.
This is 
in analogy with the Jahn-Teller theorem
in which 
lattice distortions should break crystalline symmetries
to lift degeneracy in electronic states.
\cite{ryu02-02}

\section{Coupling with lattice distortions}

As a candidate of interactions
that might lift the degeneracy of ZES's,
let us consider coupling between
the $\pi$ electrons and lattice distortions.
We use the following simple model
which is an extension of the SSH model to 2D graphite
\cite{fujita97}:
\begin{eqnarray}
H_{\mathrm{SSH}}=
\sum_{\left< ij\right>\sigma}
\big( -1+y_{ij} \big)
\big(
c_{i\sigma}^{\dagger}c_{j\sigma}
+
\mathrm{h.c.} 
\big)
+
\frac{K}{2}\sum_{\left<ij\right>} y_{ij}^{2},
\nonumber \\
\end{eqnarray}
where $c_{i\sigma}^{\dagger}/c_{i\sigma}$ is an electron creation/annihilation 
operator at a site $i$ with spin $\sigma$.
Here we treat $\sigma$-bonds as a classical spring with the tension $K$
and $y_{ij}$ represents the modulation of the length of a bond 
with a constraint $\sum_{\left<ij\right>}y_{ij}=0$.
We numerically determined $\{ y_{ij} \}$ that minimizes the ground state energy,
assuming the periodicity in $y$-direction with the unit cell shown in Fig. 
(\ref{lattice}).
Super structures with a larger unit cell 
can be energetically more favorable
\cite{harigaya92,harigaya93},
but the present choice is enough 
to demonstrate the relationship between
chiral symmetry and ZES's.

When there are no boundaries and hence no edge modes,
a first order transition with lattice distortions occurs
at $K=K_{c}\sim 1.4$ as we decrease $K$.
(Since the DOS is vanishing at the Fermi level,
the transition occurs at a finite $K$.)
When there are zigzag edges, 
one naively expects 
that the degeneracy between ZES's 
enhances Peierls instability
and a lattice distortion near edges occurs 
at some critical value of 
$K=K^{\mathrm{edge}}_{c}$ ($<K_{c}$).
However,
this is not possible as chiral symmetry is 
respected for any configuration of $\{ y_{ij} \}$
in the above SSH model.
Indeed, ZES's are stable with respect 
to lattice distortions
as shown in Fig. (\ref{spec}).

For $K<K_{c}$, 
the bulk band structure is completely reconstructed
by lattice distortions.
The corresponding loops in a parameter space are trivial
and there is no ZES 
even when one truncates the system by zigzag edges
[Fig. (\ref{spec})].

\section{Correlation effects and spin polarization}

\subsection{Mean field theory for 2D geometry}

Another source of an instability,
when there is a flat band formed by ZES's,
is electron correlation, which can give rise 
to several kinds of orders,
a typical example of which is the spin density wave.
To model the correlation effects on edge states,
let us consider the Hubbard model
defined on the honeycomb lattice with zigzag edges
\cite{fujita96}
\begin{eqnarray}
H=-
\sum_{\left<ij\right>\sigma}
\big(
c_{i\sigma}^{\dagger}c_{j\sigma}
+
\mathrm{h.c.} 
\big)
+U\sum_{i}
n_{i\uparrow}n_{i\downarrow}
-\mu
\sum_{i\sigma}n_{i\sigma}.
\nonumber \\
\end{eqnarray}
We adopt the following mean field ansatz
\begin{eqnarray}
&&
H_{\mathrm{SDW}}=
-
\sum_{\left<ij\right>\sigma}
\big(
c_{i\sigma}^{\dagger}c_{j\sigma}
+
\mathrm{h.c.} 
\big)
-\mu
\sum_{i\sigma}n_{i\sigma}
\nonumber \\
&&
-U
\sum_{i}
\left< n_{i\uparrow} \right>
\left< n_{i\downarrow}\right>
+U\sum_{i}
\big(
\left< n_{i\uparrow}\right>
n_{i\downarrow}
+
\left< n_{i\downarrow}\right>
n_{i\uparrow}
\big)
\end{eqnarray}
to investigate
possible magnetic structures,
where 
$\left< n_{i\sigma} \right>$
is determined self-consistently.
If there is no boundary,
the antiferromagnetic order occurs at $U=U_{c}\neq 0$
(Again,  the vanishing DOS at the Fermi energy 
allows the transition to occur at a finite $U$.).
$U_{c}$ is estimated as $2.23$ in MFT
whereas $U_{c}\sim 4.5$ is obtained by the quantum Monte Carlo
simulation
\cite{sorella92}.

\begin{figure}
\begin{center}
\includegraphics[width=7cm,clip]{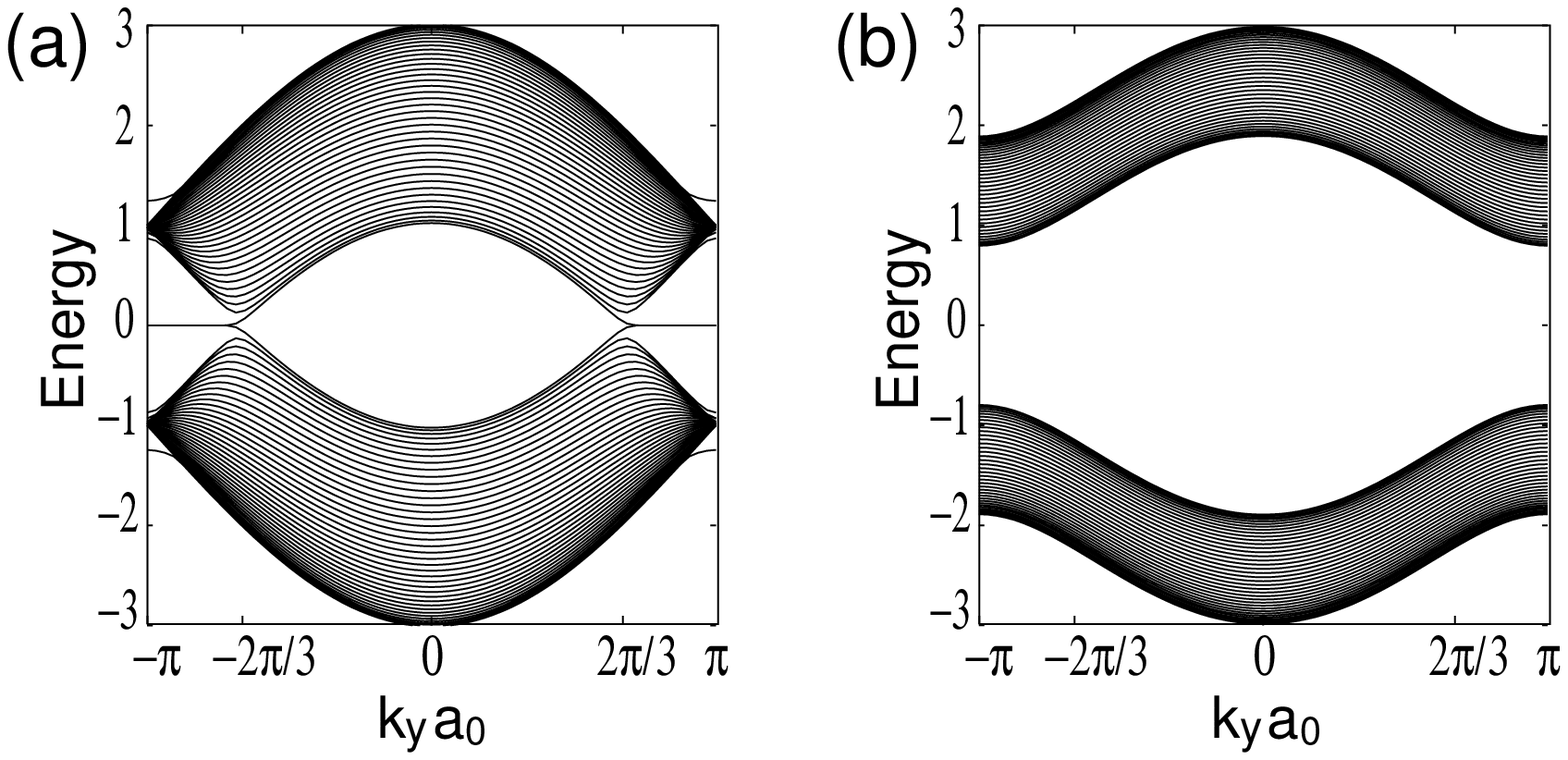}\\
\includegraphics[width=7cm,clip]{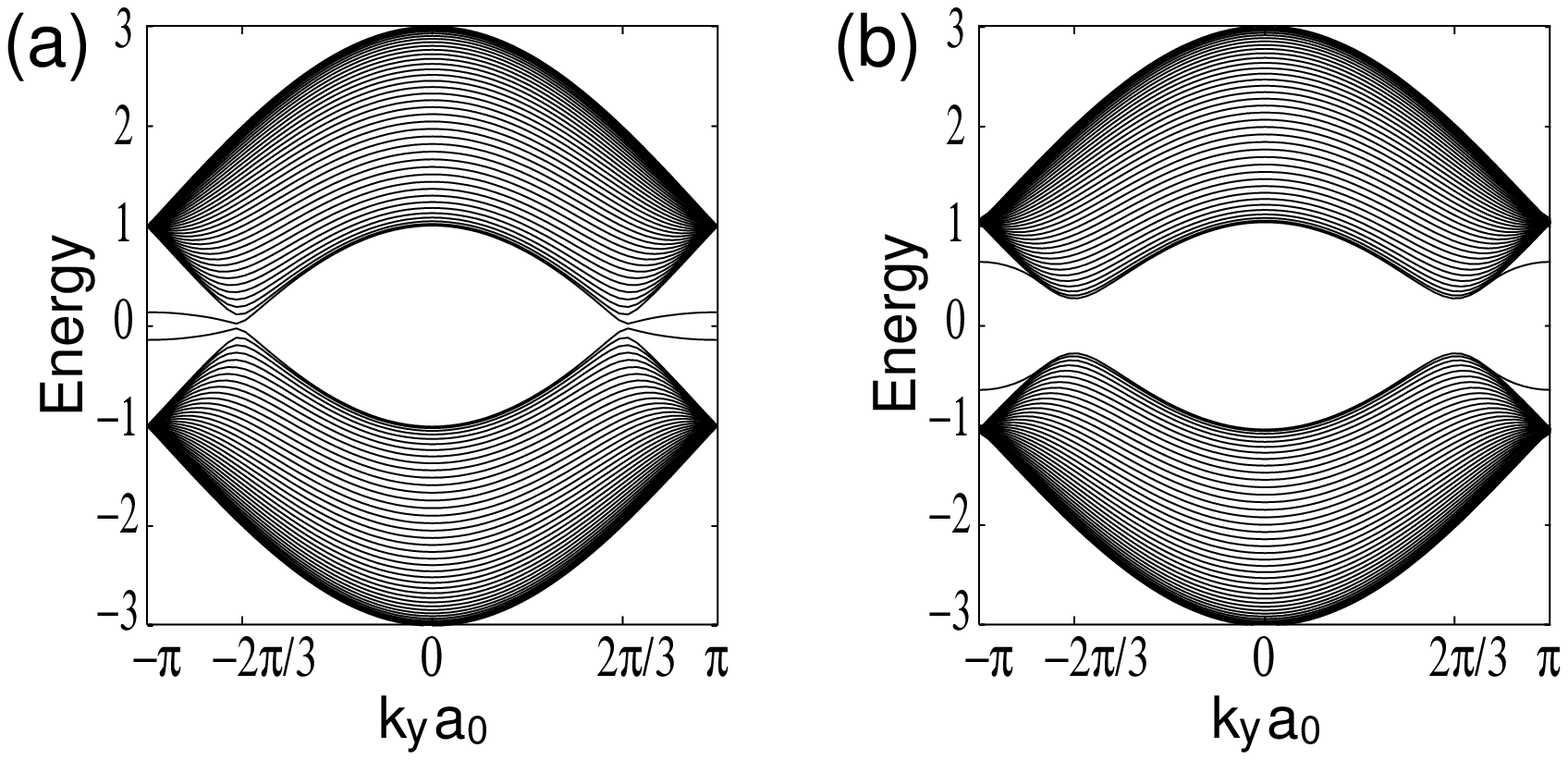}
\caption{
\label{spec}
Top:
Band structures of $\pi$-electrons 
with lattice distortions for
(a) $K=1.5 > K_{c}$ and (b) $K=1.2 < K_{c}$. 
Bottom:
Band structures 
in the background of SDW mean field
(a) $U=1.5 < U_{c}$ and (b) $U=2.5 < U_{c}$. 
}
\end{center}
\end{figure}

In contrast to lattice distortions
in the previous chapter,
a finite $\left< n_{i\sigma}\right>$
serves as an on-site potential to
break chiral symmetry
and hence lifts the degeneracy of ZES's.
Indeed,  Fujita et al.\cite{fujita96}
found that magnetic moments are induced near boundaries 
even when $U<U_{c}$ within MFT.
The calculated energy spectrum with 
the self-consistently determined mean 
field background $\left< n_{i\sigma} \right>$
exhibits the splitting 
of branches of edge modes as shown in Fig. (\ref{spec}).

\subsection{Quasi 1D geometry: carbon nanotube}

As we are discussing 2D systems with 1D edges,
MFT is not reliable enough.
Also, once we wrap a graphite sheet into 
quasi 1D tube geometry,
i.e., CNT,
the validity of MFT
is much less obvious due to the quantum fluctuation
enhanced by the low dimensionality.
We are now to give detailed analyses
for this geometry
by means of the renormalization group (RG) and
the open boundary bosonization.
\cite{ryu03}
Another quasi-1D geometry,
which is a ribbon,
was investigated by other authors.
\cite{lee02,hikihara03}
Our main focuses are on
(i) whether 
the localized charge and spin degree of freedom carried by 
ZES's can escape through the contact between
the bulk conduction electrons
and 
(ii) whether the total spin at the boundary is polarized.
The first question
 can be thought as a variant of the problem of 
quantum dissipation,
examples of which include
the Caldeira-Legget model,
the X-ray edge problem,
the Kondo effect, etc.

We consider a thin metallic $(N,-N)$ CNT,
where there exist one ($N=6$) or two ($N=9$) ZES's
when truncated with a zigzag edge. 
Starting from lattice models
with the Coulomb or the Hubbard interaction,
we can establish low-energy effective theories
which describe correlation effects at boundaries.
For the case of only one edge state,
it is given by
\begin{eqnarray}
H&=&H_{0}+H_{I},
\nonumber \\
H_{0}&=&
H_{c}+\epsilon_{e}\rho_{e}
+U_{e}n_{\uparrow}n_{\downarrow},
\nonumber \\
H_{I}&=&
\frac{v_{F}\lambda_{\rho}}{4}\,\rho_{e} \, \rho 
\nonumber \\
&&
+v_{F}\lambda_{z}\, S^{z}\, J^{z}
+\frac{v_{F}\lambda_{\perp}}{2}
\Big[
S^{+}J^{-}+J^{+}S^{-}
\Big],
\end{eqnarray}
where
$H_{c}$ is the Hamiltonian for the conduction electrons 
that includes the forward scattering part of the Coulomb interaction
by the open boundary bosonization
(small back scatterings are neglected),
$\boldsymbol{J}, \boldsymbol{S}$
represent the spin operator of the conduction electrons
and ZES's, respectively,
$\rho, \rho_{e}$
the density operator of the conduction electrons and ZES's,
and $n_{\sigma}$ denotes
the number operator for ZES's with spin $\sigma$.
$v_{F}$ represents the Fermi velocity
and an ultraviolet cutoff $\tau_{c}$
is introduced,
which is of order of the inverse bandwidth.
$U_{e}$ is the electron correlation between ZES's,
$\epsilon_{e}$ the chemical potential for ZES's,
and 
$\lambda_{\rho,z,\perp}$ represents a dimensionless coupling
between the edge states and conduction electrons.
Bare values of the Kondo-like couplings 
$\lambda_{z,\perp}$
(initial condition for RG analyses)
are ferromagnetic due to the Fermi statistics;
spin polarized states do not feel the Coulomb repulsion
as the double occupancy is automatically prohibited due to the Pauli principle,
and the kinetic energy is quenched for ZES's (Hund's law).

We perform a perturbative RG analysis up to one-loop order
by infinitesimally rescaling the ultraviolet cutoff,
$\tau_{c}\rightarrow \tau_{c}e^{-dl}$.
From the resultant RG equations, 
it can be seen
the bulk conduction electrons 
and edge states are completely decoupled in the infra-red
($l \to +\infty$),
$\lambda_{z,\perp}\to 0$.
Furthermore, 
the repulsive bulk interactions are responsible to
suppress the charge fluctuations at boundaries,
and hence
doubly occupying a edge state
becomes unfavorable in the infra-red near half filling,
$\varepsilon\to -\infty,
U_{e}+2\epsilon_{e}\to +\infty$.
Then, we conclude that 
edge states do not diffuse into the bulk
through the coupling with conduction electrons.
Our RG analysis further indicates
that a small deviation of $\epsilon_{e}$ from
the half-filling  ($\epsilon_{e}=0$) is irrelevant
due to renormalizations 
by $\lambda_{\rho},\lambda_{z,\perp}$.

In order to see
whether edge states tends to 
form a localized moments,
we need to consider 
a thicker metallic CNT which supports
more than two edge states.
We then consider 
$(9,-9)$ CNT with zigzag edges for which 
edge states appear 
for $k_{y}a_{0}=-8\pi/9$ and $+8\pi/9$.
The effective Hamiltonian 
for this case
is given by
\begin{eqnarray}
H_{0}&=&
H_{c}
+\frac{I}{4}\rho_{1}\rho_{2}
+K_{z}S_{1}^{z}S_{2}^{z}
+\frac{K_{\perp}}{2}
\Big[
S_{1}^{+}S_{2}^{-}+S_{2}^{+}S_{1}^{-}
\Big]
\nonumber \\
&&
+
U_{e}\Big[
n_{1\uparrow}n_{1\downarrow}
+n_{2\uparrow}n_{2\downarrow}
\Big]
+\epsilon_{e}\rho_{e},
\nonumber \\
H_{I}
&=&
\frac{v_{F}\lambda_{\rho}}{4}\rho_{e}\, \rho
\nonumber \\
&&
+
v_{F}\lambda_{z}S^{z}J^{z}
+
\frac{v_{F}\lambda_{\perp}}{2}
\Big[S^{+}J^{-}+J^{+}S^{-}\Big],
\end{eqnarray}
where 
$\boldsymbol{S}_{\alpha}$
represents the spin operator for ZES's
with $k_{y}a_{0}=+8\pi/9$ ($\alpha=1$)
and $k_{y}a_{0}=-8\pi/9$ ($\alpha=2$),
$\rho_{\alpha}$ the density operator,
$n_{\sigma \alpha}$ the number operator,
and 
$
\boldsymbol{S}=\boldsymbol{S}_{1}+\boldsymbol{S}_{2},
\rho_{e}=\rho_{1}+\rho_{2}
$.
Initial conditions are given by
$K_{z}=K_{\perp}=-I=-2U_{e}<0$,
$\epsilon_{e}\sim 0$,
and
$\lambda_{z}=\lambda_{\perp}\sim -\lambda_{\rho}<0$.
RG equations for $\lambda_{z}$,
$\lambda_{\perp}$,
$h_{\epsilon}$ and $h_{U}$
are identical to those in the case of one edge state.
Then, $\lambda_{z}$ and $\lambda_{\perp}$
become vanishing in the infra-red limit,
and
charge fluctuations are suppressed
$\varepsilon\to -\infty,
U_{e}+2\epsilon_{e}\to +\infty$.
On the other hand,
RG equations for $K_{z,\perp}$ and $I$,
which determine the total spin 
carried by the ground state of ZES's,
are 
\begin{eqnarray}
&&
\frac{d h_{I}}{dl}
=
h_{I}
-2\lambda_{\rho}^{2},
\nonumber \\
&&
\frac{d h_{K_{z}}}{dl}
=
h_{K_{z}}
-
\lambda_{z}^{2},\quad
\frac{d h_{K_{\perp}}}{dl}
=
h_{K_{\perp}}
-
\lambda_{\perp}^{2},
\end{eqnarray}
where $h_{K_{z,\perp}}:=\tau_{c}K_{z,\perp}$ and
$h_{I}:=\tau_{c}I$.
We see that the Kondo couplings $\lambda_{z,\perp}$ 
renormalize the
exchange interactions among edge states $K_{z,\perp}$,
making it ferromagnetic.
That is,
the couplings between bulk and edge states
turn out to 
assist the spin polarization,
leading to 
the ground state
with a highest spin $S=1$  at the boundary.

The result obtained here is consistent with 
spin polarization found in 
the mean-field theory in the previous section,
a density matrix renormalization group study
for a thin semi-conducting CNT
\cite{hikihara03},
and 
an \textit{ab initio} 
local spin density functional calculation (LSDA) \cite{okada01}
for 2D sheet geometry.

\section{Conclusions}

ZES is a hallmark of the nontrivial band of a graphite sheet,
and is related to rich physics.
We have explored several types of
order that can happen in the presence of 
electron-electron or electron-phonon interactions,
with a main emphasis on chiral symmetry and its breaking.
The Jahn-Teller-like argument presented in the present paper
can be potentially applicable to other examples.
As for spin polarization expected from chiral symmetry breaking,
we investigated its possibility for thin metallic CNT's
beyond the mean field treatment.
Our RG analysis shows that bulk part of interactions enhance
ferromagnetic interactions among edge states,
leading to spin polarization at edges. 

This work is supported by
JSPS (S.R.)
and
KAWASAKI STEEL 21st Century Foundation.

\end{document}